\documentclass[aps,pre,reprint,raggedbottom,superscriptaddress,amsfonts,amssymb,amsmath]{revtex4-1}
\usepackage{graphicx}
\usepackage[colorlinks,linkcolor=blue,citecolor=blue,urlcolor=blue]{hyperref}
\usepackage[caption=false]{subfig}
\begin{document}
\title{Onsager's Cross Coupling Effects in Gas Flows Confined to Micro-channels}
\author{Ruijie Wang}
\affiliation{School of Power and Energy, Northwestern Polytechnical University,
Xi'an 710072, Shaanxi, P.R.China}
\affiliation{Department of Mathematics, Hong Kong University of Science and
Technology, Clear Water Bay, Kowloon, Hong Kong}
\author{Xinpeng Xu}
\affiliation{Department of Mathematics, Hong Kong University of Science and
Technology, Clear Water Bay, Kowloon, Hong Kong}
\author{Kun Xu}
\affiliation{Department of Mathematics, Hong Kong University of Science and
Technology, Clear Water Bay, Kowloon, Hong Kong}
\author{Tiezheng Qian}
\thanks{Corresponding author}
\email[Email: ]{maqian@ust.hk}
\affiliation{Department of Mathematics, Hong Kong University of Science and
Technology, Clear Water Bay, Kowloon, Hong Kong}
\begin{abstract}

In rarefied gases, mass and heat transport processes interfere with each other,
leading to the mechano-caloric effect and thermo-osmotic effect,
which are of interest to both theoretical study and practical applications.
We employ the unified gas-kinetic scheme to investigate these cross coupling effects
in gas flows in micro-channels. Our numerical simulations cover
channels of planar surfaces and also channels of ratchet surfaces,
with Onsager's reciprocal relation verified for both cases.
For channels of planar surfaces, simulations are performed
in a wide range of Knudsen number and our numerical results
show good agreement with the literature results.
For channels of ratchet surfaces, simulations are performed for
both the slip and transition regimes and our numerical results not only
confirm the theoretical prediction [Phys. Rev. Lett. 107, 164502 (2011)]
for Knudsen number in the slip regime but also show that
the off-diagonal kinetic coefficients for cross coupling effects
are maximized at a Knudsen number in the transition regime.
Finally, a preliminary optimization study is carried out for the geometry
of Knudsen pump based on channels of ratchet surfaces.

\end{abstract}
\date{\today}
\maketitle

\section{Introduction}

Onsager's reciprocal relations for linear irreversible processes
\cite{Onsager1931,Onsager1931a} play a crucial and important role in the theory
of non-equilibrium thermodynamics. The kinetics of gases can be described by
elementary processes of molecular collisions where microscopic reversibility and
detailed balance are preserved. For systems slightly deviating from equilibrium,
the linear response theory applies and Onsager's reciprocal relations can be
derived with the regression hypothesis \cite{Onsager1931}. Typical examples in
gas flows are the cross coupling between mass and heat diffusion in a
multi-component gas \cite{Kremer2010}, and the mechano-caloric effect and
thermo-osmotic effect in a single-component gas \cite{Waldmann1967}. The latter
one not only provides an interesting case for theoretical study but also has
practical applications in micro-devices \cite{Sone2002, Aoki2006}.

Waldmann \cite{Waldmann1967} and Groot and Mazur \cite{Groot1984} studied the
cross coupling effect in channels of parallel planar surfaces in both
the free molecular (${\rm Kn}\ge 10$) and slip ($0.001\le {\rm Kn}\le 0.1$) regimes.
Loyalka \cite{Loyalka1971,Loyalka1975} and Sharipov \cite{Sharipov2006} analyzed
the cross coupling effect by means of the linearized Boltzmann method and
obtained the coupling coefficients numerically. Although the theoretical
analysis of Loyalka \cite{Loyalka1971} is valid for capillaries of arbitrary
shape, most of the work, especially the numerical calculation
\cite{Loyalka1975,Sharipov1998}, was devoted to capillaries of planar surfaces
or circular cross sections. The thermo-osmotic effect has attracted much
attention recently since it can be used to design pumping devices without any
moving part, i.e. the Knudsen pump \cite{Sone2002}. In addition to the earlier
proposed Knudsen pump \cite{Sone2002,Aoki2006}, capillaries with ratchet
surfaces have the potential for other possible configurations \cite{Wurger2011}.
The driving mechanism of these systems has been analyzed by W\"uger
\cite{Wurger2011} as well as Hardt et al. \cite{Hardt2013}, and the mass and
momentum transfer has been studied by Donkov et al. \cite{Donkov2011}.

In the present work, we will study the cross coupling phenomena for a long
capillary by using the unified gas-kinetic scheme \cite{Xu2010,Huang2012}. We
will study the case of planar surfaces as well as the case of ratchet surfaces.
The cross coupling mechanism will be presented for both cases. The coupling
coefficients for the case of planar surfaces are numerically calculated and
compared to literature results. The coupling coefficients for the case of
ratchet surfaces are numerically calculated and analyzed, with a comparison to
literature results as well. A preliminary geometry optimization for the design
of Knudsen pump is also presented.

In our simulations for the case of ratchet surfaces, a channel of finite length
is used with the pressure boundary condition applied at the two ends. In the
presence of both pressure difference and temperature difference, the cross
coupling effects, namely the mechano-caloric effect and thermo-osmotic effect,
can be jointly studied to reveal the underlying reciprocal symmetry. However,
with the periodic boundary condition used by Donkov et al.\cite{Donkov2011},
only the thermo-osmotic effect is attainable because no pressure difference can
be applied to generate the mechano-caloric effect. The numerical scheme used in
the present work is applicable for all Knudsen numbers, and this allows us to
study the cross coupling phenomena beyond the slip regime in which W\"uger's
theory \cite{Wurger2011} is valid. In particular, we find that the cross
coupling effects are maximized at a Knudsen number in the transition regime.

\section{Cross Coupling in Gas Flows in Micro-channels}

In a closed system out of equilibrium, the rate of entropy production can be
expressed as
\begin{equation}
    \frac{dS}{dt}=\sum_{i=1}^NJ_iX_i,
\end{equation}
where $S$ is the entropy, $J_i$ are the thermodynamic fluxes, and $X_i$ are the
conjugate thermodynamic forces. For small deviation away from equilibrium, we
have the linear relations between $J_i$ and $X_i$:
\begin{equation}
    J_i=\sum_{j=1}^NL_{ij}X_j,
\end{equation}
where $L_{ij}$ are the kinetic coefficients. Onsager's reciprocal relations
state that $L_{ij}$ and $L_{ji}$ are equal as a result of microscopic
reversibility \cite{Onsager1931, Onsager1931a,gyarmati1970}. Starting from the
Gibbs equation, the thermodynamic fluxes and forces can be identified for gas
flows, and the corresponding constitutive equations can be derived
\cite{Kremer2010}.

\subsection{Micro-channels of planar surfaces}

\begin{figure}
    \centering
    \includegraphics[width=\columnwidth]{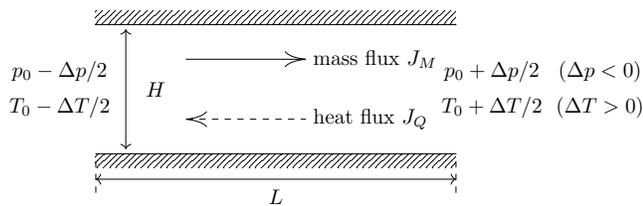}
    \caption{A schematic illustration of the cross coupling in a channel of
    planar surfaces.}
    \label{fg:schematic_coupling_planar}
\end{figure}

A schematic illustration of the cross coupling in a channel of planar surfaces
can be found in figure \ref{fg:schematic_coupling_planar}, where a long channel
is confined by two parallel solid plates separated by a distance $H$ and
connected with two reservoirs. The left reservoir is maintained at pressure
$p_0-\Delta p/2$ and temperature $T_0-\Delta T/2$ while the right reservoir is
maintained at $p_0+\Delta p/2$ and $T_0+\Delta T/2$. We use $\Delta p <0$ and
$\Delta T >0$ in our simulations, with $|\Delta p/p_0|\ll 1$ and $|\Delta
T/T_0|\ll 1$ to ensure the linear response. Usually, a mass flux to the right is
generated by the pressure gradient due to $\Delta p <0$ and a heat flux to the
left is generated by the temperature gradient due to $\Delta T >0$. For rarefied
gas, however, $\Delta p$ also contributes to the heat flux and $\Delta T$ also
contributes to the mass flux. These cross coupling effects are called the
mechano-caloric effect and thermo-osmotic effect respectively.

For a single-component gas, the rate of entropy production can be expressed as
\cite{Waldmann1967}
\begin{equation}
    \frac{dS}{dt}=
    J_M\Delta\left(-\frac{\nu}{T}\right)+J_E\Delta\left(\frac{1}{T}\right),
    \label{eq:entropy_production}
\end{equation}
where $\nu$ is the chemical potential per unit mass, $J_E$ and $J_M$ are the
energy flux and mass flux from the left reservoir to the right reservoir, and
$\Delta$ means the quantity on the right minus the quantity on the left. Here
$\nu$ and $J_E$ can be written as
\begin{gather}
    \nu=h - Ts,\\
    J_E=J_Q+hJ_M,
\end{gather}
where $s$ and $h$ are the entropy and enthalpy per unit mass, and $J_Q$ is the
heat flux. Together with the Gibbs-Duhem equation
\begin{equation}
    d\nu=-sdT+dp/\rho,
\end{equation}
where $\rho$ is the mass density. Equation \eqref{eq:entropy_production} becomes
\begin{equation}
    \frac{dS}{dt}=
    -\frac{1}{\rho T}J_M \Delta p
    -\frac{1}{T^2}J_Q \Delta T.
    \label{eq:entropy_production_transformed}
\end{equation}

According to equation \eqref{eq:entropy_production_transformed}, the
thermodynamic forces and fluxes are connected in the form of
\begin{equation}
    \begin{bmatrix}J_M\\ J_Q\end{bmatrix} =
    \begin{bmatrix}L_{MM} & L_{MQ}\\ L_{QM} & L_{QQ}\end{bmatrix}
    \begin{bmatrix}-\rho_0^{-1}T_0^{-1}\Delta p\\
    -T_0^{-2}\Delta T\end{bmatrix},
    \label{eq:force-flux-matrix}
\end{equation}
with
\begin{equation}
    L_{MQ}=L_{QM},
\end{equation}
due to Onsager's reciprocal relations. The detailed mechanism may vary with
geometric configuration and rarefaction. Here and throughout the paper,
the subscript `0' denotes the reference state from which
various deviations (in pressure, temperature, etc) are measured.

In the free molecular regime and with specular reflection on plates, the gas
molecules travel ballistically from on side to the other and the distribution
function at any point can be treated as a combination of two half-space
maxwellians from the two reservoirs. The kinetic coefficients in equation
\eqref{eq:force-flux-matrix} can be analytically derived in this case
\cite{Waldmann1967}, given by
\begin{equation}
    \begin{aligned}
        \begin{bmatrix}L_{MM} & L_{MQ}\\ L_{QM} & L_{QQ}\end{bmatrix}&\\
        =
        \frac{H\rho_0 T_0}{4}&\sqrt{\frac{8k_BT_0}{\pi m}}
        \begin{bmatrix}
            \rho_0/p_0 & -1/2\\[5pt] -1/2 & 9p_0/4\rho_0
        \end{bmatrix}.
    \end{aligned}
\end{equation}
where $k_B$ is the Boltzmann constant and $m$ is the molecular mass.

If the temperature gradient is imposed on the plates and the gas molecules are
diffusely reflected, then the mass flux due to the temperature gradient is
generated by thermal creep on the plates \cite{Groot1984,Sone2002,Shen2005}. The
kinetic coefficients in this case have been calculated by several authors using
different methods \cite{Sharipov1998}. Assuming the length to height ratio of
the channel is fixed and noting $\rho\lambda={\rm constant}$ and $\mu\propto
\rho^0 T^{1/2}$ for hard-sphere molecules, the average velocity $\overline{U}$
induced by thermal creep can be estimated from the Maxwell slip boundary
condition \cite{Shen2005},
\begin{equation}
    \overline{U}\sim\frac{\mu_0}{\rho_0 T_0}\nabla T\propto
    {\rm Kn}\frac{\Delta T}{\sqrt{T_0}},
    \label{eq:estimate_velocity_planar}
\end{equation}
where $\lambda$ is the mean free path, $\mu$ is the dynamic viscosity
independent of the density,
and ${\rm Kn}=\lambda_0/H$ is the Knudsen number. In later sections, we will show
that $L_{MQ}$ and $L_{QM}$ are equal and increase with the increasing ${\rm Kn}$.

\subsection{Micro-channels of ratchet surfaces}

\begin{figure*}
    \centering
    \subfloat[]{
        \includegraphics[width=0.55\linewidth]{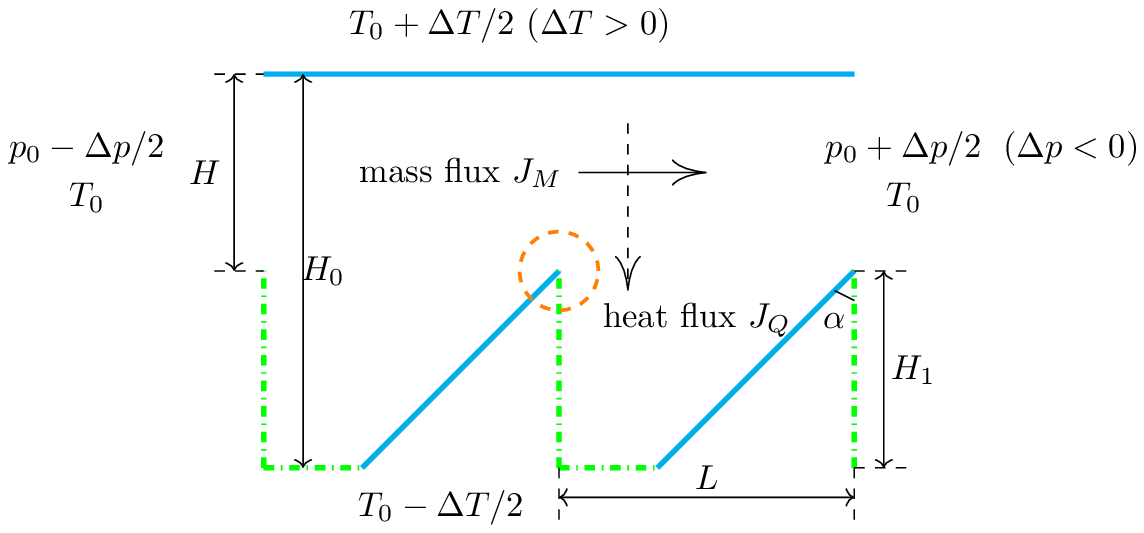}
        \label{fg:schematic_coupling_ratchet_less}
    }
    \vfill
    \subfloat[]{
        \includegraphics[width=0.45\linewidth]{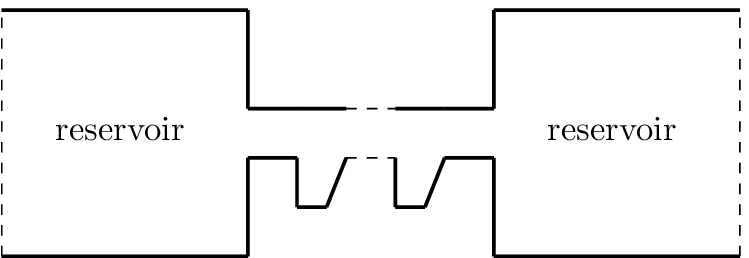}
        \label{fg:schematic_coupling_ratchet_more}
    }
    \caption{A schematic illustration of the cross coupling in a channel of
        ratchet surfaces.
        (a) Two blocks of a channel of ratchet surfaces. The upper wall and the
        lower tilted walls (blue solid lines) are diffusely reflective. The
        lower horizontal and vertical walls (green dot-dashed lines) are
        specularly reflective.
        (b) Simulation geometry with multiple blocks in the channel.
    }
    \label{fg:schematic_coupling_ratchet}
\end{figure*}

The mechanism for the cross coupling in a channel of ratchet surfaces is more
complicated. Consider a long channel consisting of repeating structure
\cite{Donkov2011} as shown in figure \ref{fg:schematic_coupling_ratchet_less},
where the two ends are connected to two reservoirs maintained at $(p_0-\Delta
p/2, T_0)$ on the left and $(p_0+\Delta p/2,T_0)$ on the right with $\Delta
p<0$. The upper wall (blue solid lines) is diffusely reflective and maintained
at $T_0+\Delta T/2$ with $\Delta T>0$, and the lower tilted walls (blue solid
lines) are diffusely reflective and maintained at $T_0-\Delta T/2$. The lower
horizontal and vertical walls (green dot-dashed lines) are specularly
reflective. Usually, a mass flux from the left to the right is generated by the
pressure gradient due to $\Delta p<0$, and a heat flux from the top to the
bottom is generated by the temperature gradient due to $\Delta T>0$. For
rarefied gas, however, $\Delta p$ also contributes to the vertical heat flux and
$\Delta T$ also contributes to the horizontal mass flux. Our simulations are
carried out for $|\Delta p/p_0|\ll 1$ and $|\Delta T/T_0|\ll 1$ to ensure the
linear response.

In general, a temperature difference can also be applied in the horizontal
direction and a heat flux can also exist in this direction as the conjugate
flux. This means that there are three thermodynamic forces (i.e. the pressure
difference in the horizontal direction, the temperature difference in the
horizontal direction, and the temperature difference in the vertical direction)
and their conjugate fluxes. In the present study, we set the temperature
difference in the horizontal direction to be zero. We also ignore the heat flux
in the horizontal direction (although it may well exist). As a result, the two
thermodynamic forces and their conjugate fluxes under consideration are
connected in the form of
\begin{equation}
    \begin{bmatrix}J_M\\ J_Q\end{bmatrix} =
    \begin{bmatrix}L_{MM} & L_{MQ}\\ L_{QM} & L_{QQ}\end{bmatrix}
    \begin{bmatrix}-\rho_0^{-1}T_0^{-1}\Delta p\\
    -T_0^{-2}\Delta T\end{bmatrix},
\end{equation}
with
\begin{equation}
    L_{MQ}=L_{QM},
\end{equation}
due to Onsager's reciprocal relations. Here $J_M$ is the {\it horizontal}
mass flux (from the left to the right) and $J_Q$ is the {\it vertical} heat flux
(from the bottom to the top).

Since the upper wall and the lower tilted walls are maintained at different
temperatures (with $\Delta T>0$), the isothermal lines near the tips of the
tilted walls (indicated by a red dashed circle) are sharply curved and a thermal
edge flow is induced at the tip from the top to the bottom \cite{Sone2002,
Hardt2013}. It is possible to make a rough estimation of the induced flow
velocity at the tip \cite{Sone2002,Wurger2011}. If all the walls are assumed to
be diffusely reflective, then the temperature gradient along the tilted wall
near the tip is approximated by
\begin{equation}
    \nabla_{||}T=\Delta T
    \frac{\pi^2}{(2\pi-\alpha)^2}\cos\left(\frac{\alpha}{2}\right)
    \frac{H^{-\pi/(2\pi-\alpha)}}{\lambda_0^{1-\pi/(2\pi-\alpha)}},
\end{equation}
where $\lambda_0$ is the mean free path at $(p_0,T_0)$ \cite{Wurger2011}. By use
of $\rho\lambda={\rm constant}$ and $\mu\propto \rho^0 T^{1/2}$ for hard-sphere
molecules, the induced velocity in the slip regime can be estimated from the
slip boundary condition \cite{Shen2005}:
\begin{equation}
    \begin{aligned}
        {\overline U} \sim & \frac{\mu_0}{\rho_0 T_0}\nabla_{||}T \\
        \propto &
        \frac{\Delta T}{\sqrt{T_0}}
        \frac{1}{(2\pi-\alpha)^2}\cos\left(\frac{\alpha}{2}\right)
        \left(\frac{\lambda_0}{H}\right)^{\pi/(2\pi-\alpha)}.
    \end{aligned}
    \label{eq:ratchet_average_velocity_assumed}
\end{equation}
For $0\le\alpha<\pi/2$ and ${\rm Kn}=\lambda_0/H<1$ in the slip regime, the
induced velocity is
\begin{enumerate}
    \item proportional to $\Delta T$,
    \item a decreasing function of $T_0$,
    \item a decreasing function of $\alpha$ because a smaller $\alpha$ means
        sharper edges,
    \item an increasing function of ${\rm Kn}$ because a larger ${\rm Kn}$ leads
        to stronger non-equilibrium effect.
\end{enumerate}

It is worth noting that $\overline{U}$ will decrease if the Knudsen number
exceeds a certain value since the thermally induced flows are typically
strongest in the lower transition regime (where the Knudsen number is slightly
higher than that in the slip regime) \cite{Sone2002}. For the current
configuration, the average velocity can be expressed as
\begin{equation}
    \overline{U}\propto {\rm Kn}^{C_2} \frac{\Delta T}{\sqrt{T_0}},
    \label{eq:estimate_velocity_ratchet}
\end{equation}
where $C_2$ is a constant determined by a specific geometry.

\section{Numerical Results and Discussion}

The kinetic coefficients are to be calculated and presented in dimensionless
form as
\begin{equation}
    \begin{aligned}
        {\hat L}_{MQ}&= - L_{MQ}\left(\frac{2k_B}{m\rho_0 C_0^3
        H}\right),\\[5pt]
        {\hat L}_{QM}&= - L_{QM}\left(\frac{2k_B}{m\rho_0 C_0^3
        H}\right),\\[5pt]
    \end{aligned}
\end{equation}
where $\rho_0$ and $T_0$ are the density and temperature of the reference state,
$C_0=\sqrt{2k_BT_0/m}$ is the most probable speed, and $H$ is the height of the
channel to define the Knudsen number ${\rm Kn}=\lambda_0/H$. The mean free path
in the reference state is $\lambda_0$. The above normalization makes it easier
to compare our results with the benchmark solutions, and the normalized
coefficients can reflect the mechanism more directly as shown below.

As the density variation is small in the simulation, the mass flux can be
expressed as
\begin{equation}
    J_{M}\approx\rho_0\overline{U}H.
\end{equation}
Assuming that there is no pressure difference, we have
\begin{equation}
    \rho_0\overline{U}H={\hat L}_{MQ}
    \left(\frac{m\rho_0 C_0^3H}{2k_B}\right)
    \frac{\Delta T}{T_0^2},
\end{equation}
which leads to
\begin{equation}
    \overline{U}={\hat L}_{MQ}C_0\frac{\Delta T}{T_0}
    \propto {\hat L}_{MQ}\frac{\Delta T}{\sqrt{T_0}}.
    \label{eq:relation_velocity_coefficients}
\end{equation}
Through a comparison of equation \eqref{eq:relation_velocity_coefficients} with
equations \eqref{eq:estimate_velocity_planar} and
\eqref{eq:estimate_velocity_ratchet}, the dimensionless ${\hat L}_{MQ}$ is
expected to have the form
\begin{equation}
    {\hat L}_{MQ}=C_1{\rm Kn}^{C_2},
    \label{eq:coefficient_assumed_form}
\end{equation}
where $C_1$ and $C_2$ are constants determined by a specific geometry. They are
to be obtained by fitting the simulation data.

For the cross coupling considered here, two thermodynamic fluxes (mass flux and
heat flux) and the corresponding forces (pressure difference and temperature
difference) can be directly extracted from the simulation data in a single
simulation. In order to determine all the kinetic coefficients, simulations need
to be performed (at least) twice with different $\Delta p$ and $\Delta T$ for
the same system (with the same geometry and the same Knudsen number).

\subsection{Cross coupling in channels of planar surfaces}

\begin{figure}
    \centering
    \includegraphics[width=0.9\columnwidth]{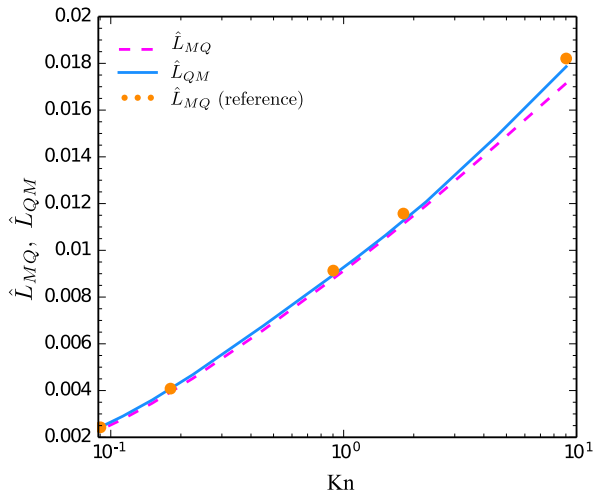}
    \caption{
        The normalized off-diagonal coefficients ${\hat L}_{MQ}$ and ${\hat L}_{QM}$
        versus the Knudsen number. The reference is the S-model solution based on
        the variational method by Chernyak et al. \cite{Chernyak1979}.
    }
    \label{fg:coupling_coefficients_planar}
\end{figure}

First we calculate the kinetic coefficients for channels with planar surfaces
and compare our results with those in literature \cite{Sharipov1998,
Chernyak1979, Cercignani1988}. A schematic illustration for the simulation
geometry can be found in figure \ref{fg:schematic_coupling_planar}. The solid
surfaces are diffusely reflective and have linearly distributed temperature from
$T_0-\Delta T/2$ to $T_0+\Delta T/2$. The gas particles are hard-sphere and
monatomic, with the Prandtl number ${\rm Pr}=2/3$ and the dynamic viscosity
$\mu\propto T^{0.5}$. The Knudsen number is defined as ${\rm Kn}=\lambda_0/H$.
$\Delta p$ and $\Delta T$ are kept small enough so that the response of fluxes
to forces is linear. The length to height ratio of the channel is taken to be
$20$ in order to reduce the influence of inlet and outlet. When extracting the
kinetic coefficients, the pressure and temperature differences are measured at
the inlet and outlet, the mass flux $J_M$ is measured over the cross section at
the inlet and outlet, and the heat flux $J_Q$ is measured over the cross section
in the middle of the channel.

Figure \ref{fg:coupling_coefficients_planar} shows the normalized off-diagonal
coefficients ${\hat L}_{MQ}$ and ${\hat L}_{QM}$ versus the Knudsen number. The
two coefficients are very close to each other with a relative difference less
than $5\%$, and have good agreement with the S-model solution based on the
variational method by Chernyak et al. \cite{Chernyak1979}. The off-diagonal
coefficients are zero at ${\rm Kn}=0$ since there is no thermally induced flow
in the continuum limit and the heat flux simply follows Fourier's law. The
normalized coefficients increase with the increasing Knudsen number. In the slip
regime ($0.001<{\rm Kn}<0.1$), ${\hat L}_{MQ}\sim{\rm Kn}$, while at large
Knudsen numbers (${\rm Kn}>3$), the profile is almost linear, which means ${\hat
L}_{MQ}={\hat L}_{QM} \sim\log({\rm Kn})$. This agrees with the conclusion
obtained from linearized Boltzmann equation for two-dimensional infinitely long
channels \cite{Cercignani1988,Sharipov1998}.

\subsection{Cross coupling in channels of ratchet surfaces}

\begin{figure*}
    \centering
    \subfloat[]{
        \includegraphics[height=0.40\linewidth]{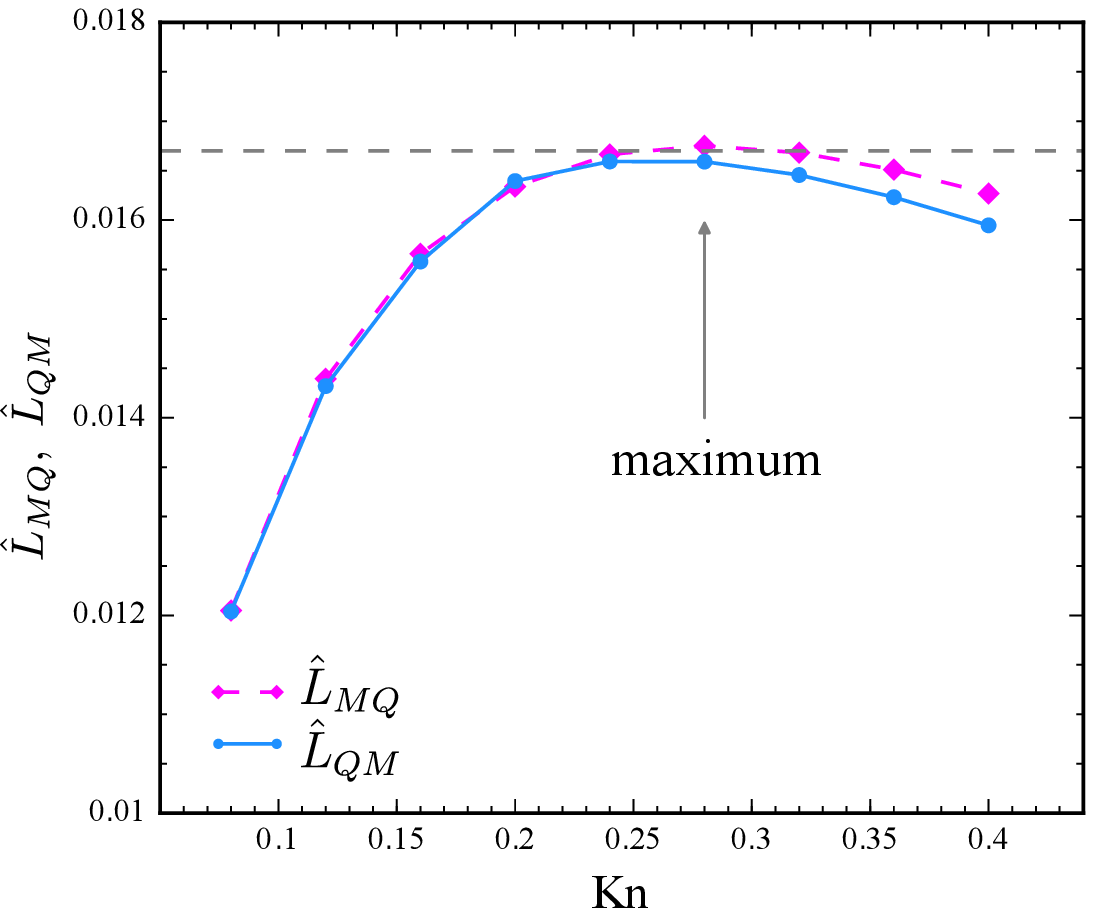}
        \label{fg:coupling_coefficients_ratchet_simulation}
    }
    \subfloat[]{
        \includegraphics[height=0.40\linewidth]{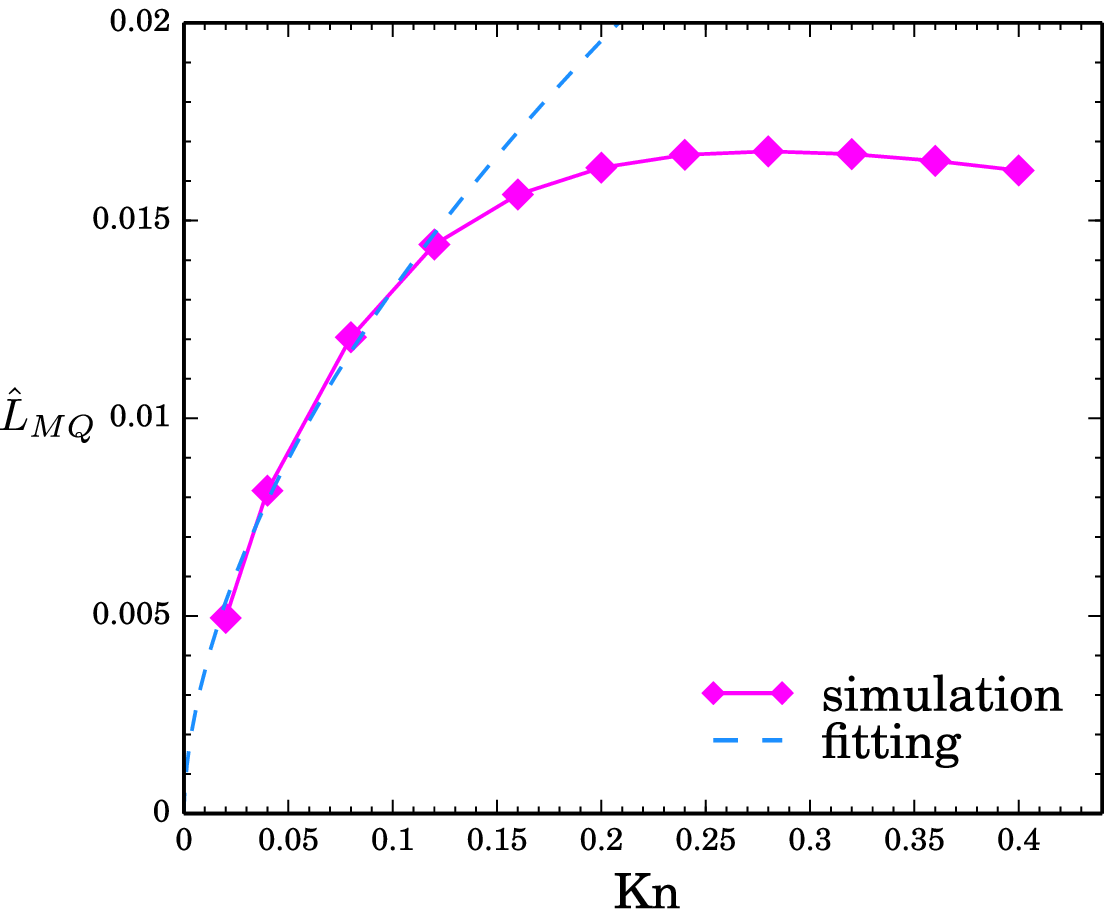}
        \label{fg:coupling_coefficients_ratchet_fitting}
    }
    \caption{
        (a) The normalized off-diagonal coefficients ${\hat L}_{MQ}$ and ${\hat
        L}_{QM}$ versus the Knudsen number. The dashed line is used to indicate
        that ${\hat L}_{MQ}$ and ${\hat L}_{QM}$ both exhibit their maxima at
        ${\rm Kn}\approx 0.28$.
        (b) ${\hat L}_{MQ}$ from simulations. For ${\rm Kn}\le 0.12$, the data
        fitting gives ${\hat L}_{MQ}=0.0483{\rm Kn}^{0.562} $, represented by
        the dashed line.
    }
    \label{fg:coupling_coefficients_ratchet}
\end{figure*}

In our simulations, each channel consists of seven repeating blocks as shown in
figure \ref{fg:schematic_coupling_ratchet_less} and each block has $L/H_0=1$,
$H_1/H=1$, and $\alpha=45^\circ$. In addition, two parallel sections with
specularly reflective walls of length $L$ are attached at the two ends. The
channel is then connected to two reservoirs. A schematic illustration for the
whole system is shown in figure \ref{fg:schematic_coupling_ratchet_more}. The
parallel sections with specularly reflective walls connected to the reservoirs
are introduced to reduce the
boundary effect of the inlet/outlet on the ratchet sections and to simplify the
measurement of mass flux. The gas particles are still hard-sphere and monatomic,
with the Prandtl number ${\rm Pr}=2/3$ and the dynamic viscosity $\mu\propto
T^{0.5}$. The Knudsen number is defined as ${\rm Kn}=\lambda_0/H$. When
extracting the kinetic coefficients, the pressure difference is measured at the
inlet and outlet, the mass flux $J_M$ is measured over the cross section at the
inlet and outlet, and the heat flux $J_Q$ is measured over all the lower tilted
walls.

Figure \ref{fg:coupling_coefficients_ratchet_simulation} shows the normalized
off-diagonal coefficients ${\hat L}_{MQ}$ and ${\hat L}_{QM}$ versus the Knudsen
number. The two coefficients are very close to each other with a relative
difference less than $2\%$, indicating that the relation $L_{MQ}=L_{QM}$ is well
satisfied in our simulations. The off-diagonal coefficients are zero at ${\rm
Kn}=0$ since there is no thermally induced flow in the continuum limit and the
heat flux simply follows Fourier's law. As the Knudsen number increases from
zero, the rarefaction effects begin to emerge at the sharp edge of the ratchet
and lead to nonzero ${\hat L}_{MQ}$ and ${\hat L}_{QM}$. It is further noted
that ${\hat L}_{MQ}$ and ${\hat L}_{QM}$ both exhibit their maxima at ${\rm
Kn}\approx 0.28$. This value of ${\rm Kn}$ (in the transition regime) is
consistent with the results for ${\hat L}_{MQ}$
in Ref. \cite{Donkov2011} for a similar ratchet geometry
with periodic boundary condition. Physically, the cross coupling
effects arise from the thermally induced flow, which is the strongest at the
sharp edge. As the Knudsen number becomes higher than a certain value, the
surface structure is no longer clearly detectable by the particles, and hence
the off-diagonal coefficients will decrease with the increasing Knudsen number.
From the distinct behaviors of ${\hat L}_{MQ}$ as a function of ${\rm Kn}$ in
the slip and transition regimes, a maximum is expected for ${\hat L}_{MQ}$.

Now we perform the data fitting according to $L_{MQ}=L_{QM}$ and equation
\eqref{eq:coefficient_assumed_form}. Since this formula is valid in the slip
regime ($0.001\le {\rm Kn}\le 0.1$),
two additional simulations are carried out to obtain ${\hat L}_{MQ}$ at
${\rm Kn}=0.02$ and $0.04$, as shown in figure
\ref{fg:coupling_coefficients_ratchet_fitting}. The data fitting uses only the
first four data points (for ${\rm Kn}\le 0.12$, see figure
\ref{fg:coupling_coefficients_ratchet_fitting}) and gives
\begin{equation}
    {\hat L}_{MQ}=0.0483{\rm Kn}^{0.562}.
    \label{eq:ratchet_coefficient_fitted}
\end{equation}
Inserting $\alpha=\pi/4$ into equation
\eqref{eq:ratchet_average_velocity_assumed}, we find
\begin{equation}
    {\bar U}\propto {\rm Kn}^{4/7},
\end{equation}
in which the exponent $4/7\simeq 0.571$ is very close to the fitting parameter
${0.562}$ in equation \eqref{eq:ratchet_coefficient_fitted}. Physically, the
exponent $C_2$ directly reflects the underlying cross coupling mechanism, and
therefore our simulation results have confirmed the theory proposed in Ref.
\cite{Wurger2011} for ${\rm Kn}$ in the slip regime.

\section{Knudsen Pump}

\begin{figure*}
    \centering
    \subfloat[]{\includegraphics[width=0.35\linewidth]{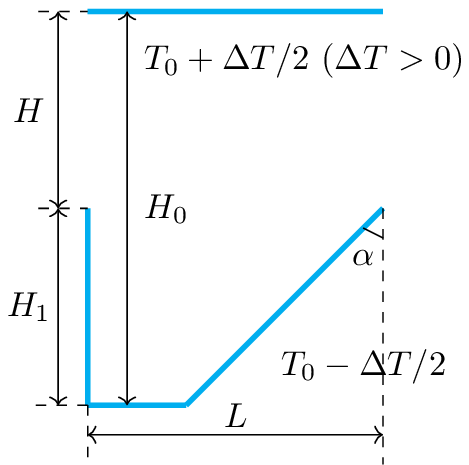}}
    \subfloat[]{\includegraphics[width=0.35\linewidth]{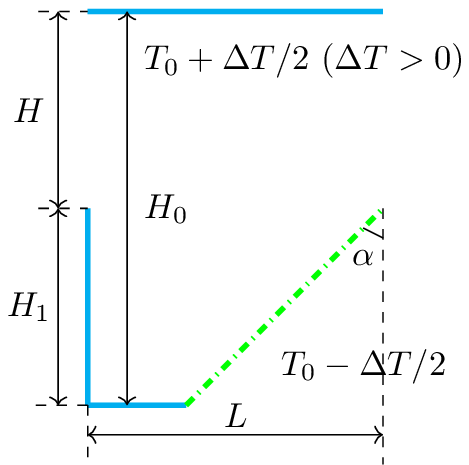}}
    \caption{
        (a) The diffuse configuration.
        (b) The diffuse-specular configuration.
    }
    \label{fg:schematic_knudsen_pump}
\end{figure*}

\begin{figure*}
    \centering
    \subfloat[]{
        \includegraphics[height=0.40\linewidth]{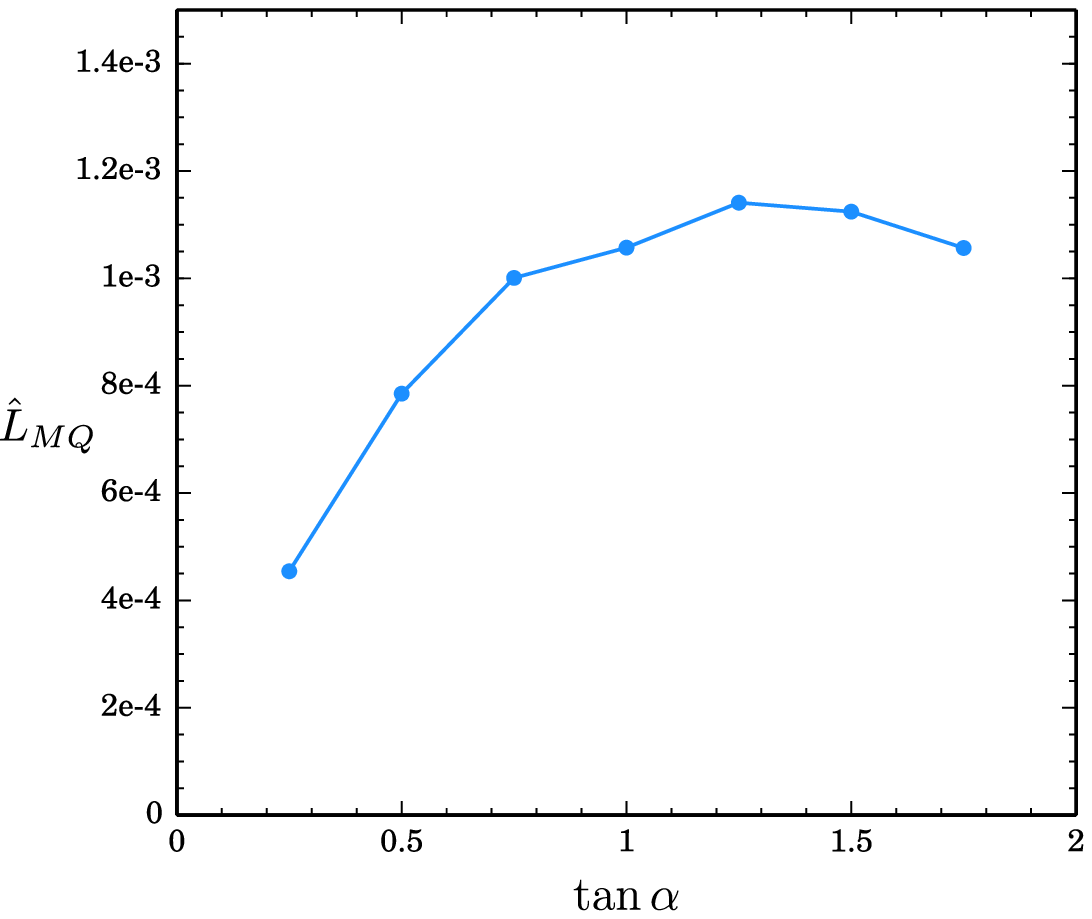}
        \label{fg:coupling_ratchet_optimize_alpha_diffuse}
    }
    \subfloat[]{
        \includegraphics[height=0.40\linewidth]{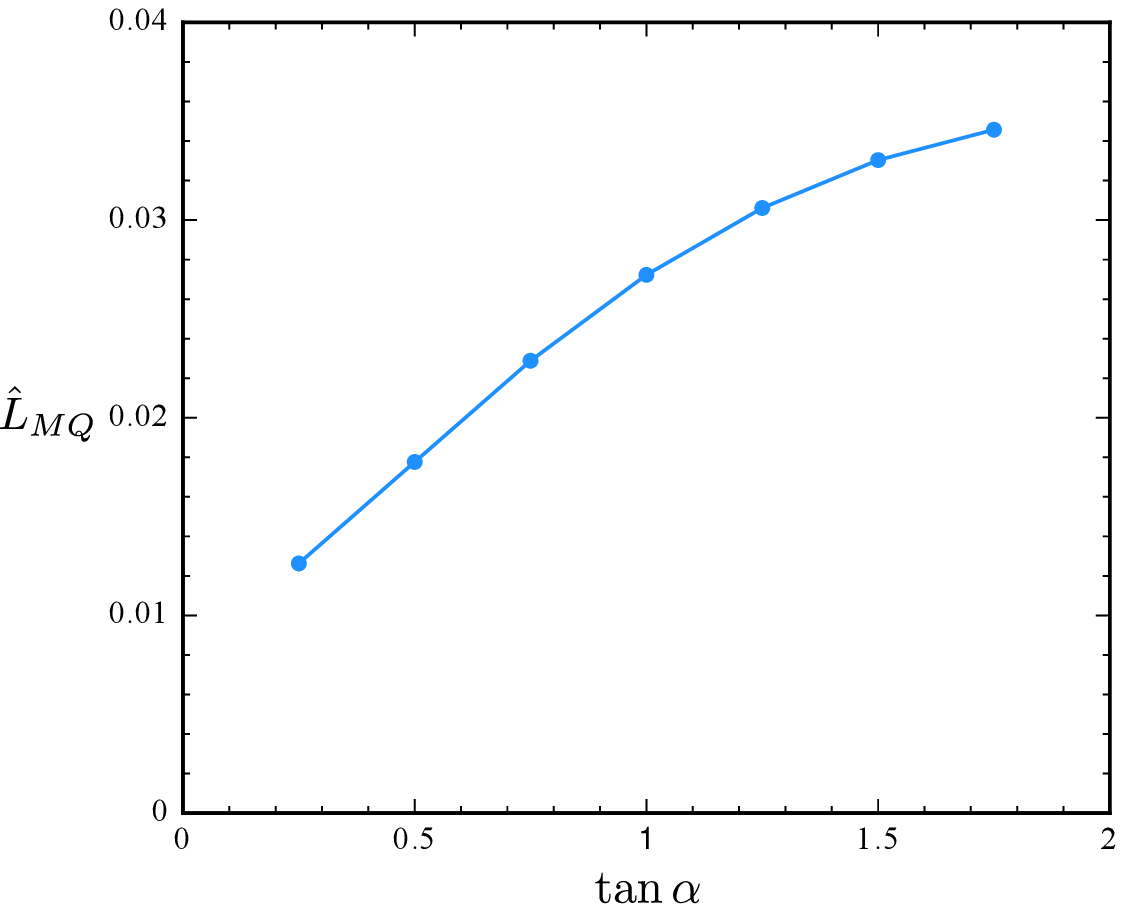}
        \label{fg:coupling_ratchet_optimize_alpha_specular}
    }
    \caption{${\hat L}_{MQ}$ plotted as a function of $\tan\alpha$ for $L/H=2$,
        $H_1/H=1$, and ${\rm Kn}=0.28$.
        (a) ${\hat L}_{MQ}$ in the diffuse configuration.
        (b) ${\hat L}_{MQ}$ in the diffuse-specular configuration.
    }
    \label{fg:coupling_ratchet_optimize_alpha}
\end{figure*}

\begin{figure*}
    \centering
    \subfloat[]{
        \includegraphics[height=0.40\linewidth]{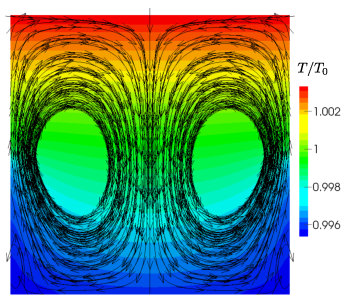}
        \label{fg:coupling_ratchet_field_diffuse}
    }
    \subfloat[]{
        \includegraphics[height=0.40\linewidth]{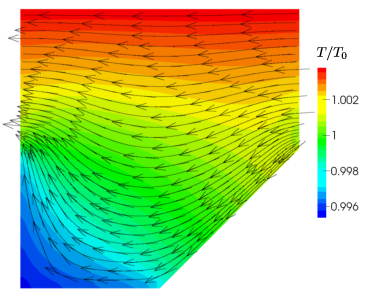}
        \label{fg:coupling_ratchet_field_specular}
    }
    \caption{The temperature distribution and streamlines plotted in one block
        with the periodic boundary condition.
        (a) The diffuse configuration with $\alpha=0$.
        (b) A typical diffuse-specular configuration.
    }
    \label{fg:coupling_ratchet_field}
\end{figure*}

\begin{figure*}
    \centering
    \subfloat[]{
        \includegraphics[height=0.40\linewidth]{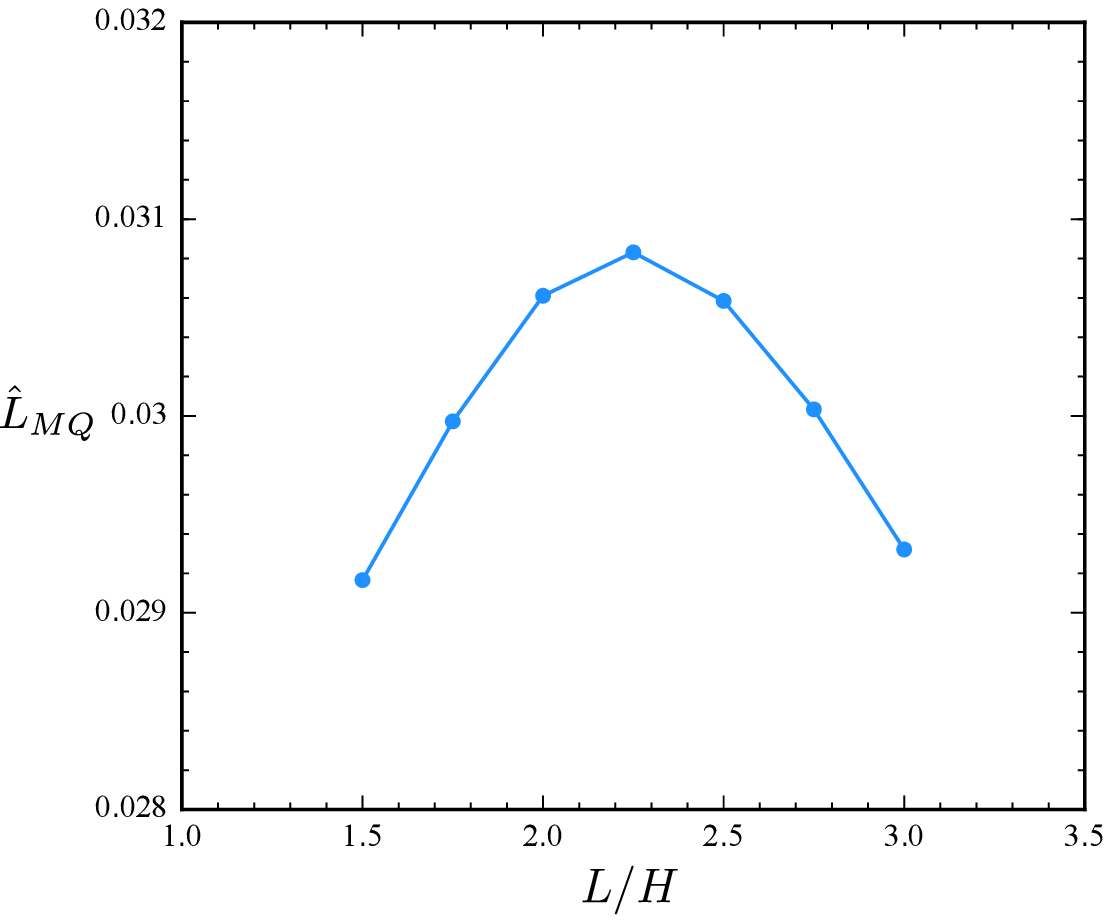}
        \label{fg:coupling_ratchet_optimize_length}
    }
    \subfloat[]{
        \includegraphics[height=0.40\linewidth]{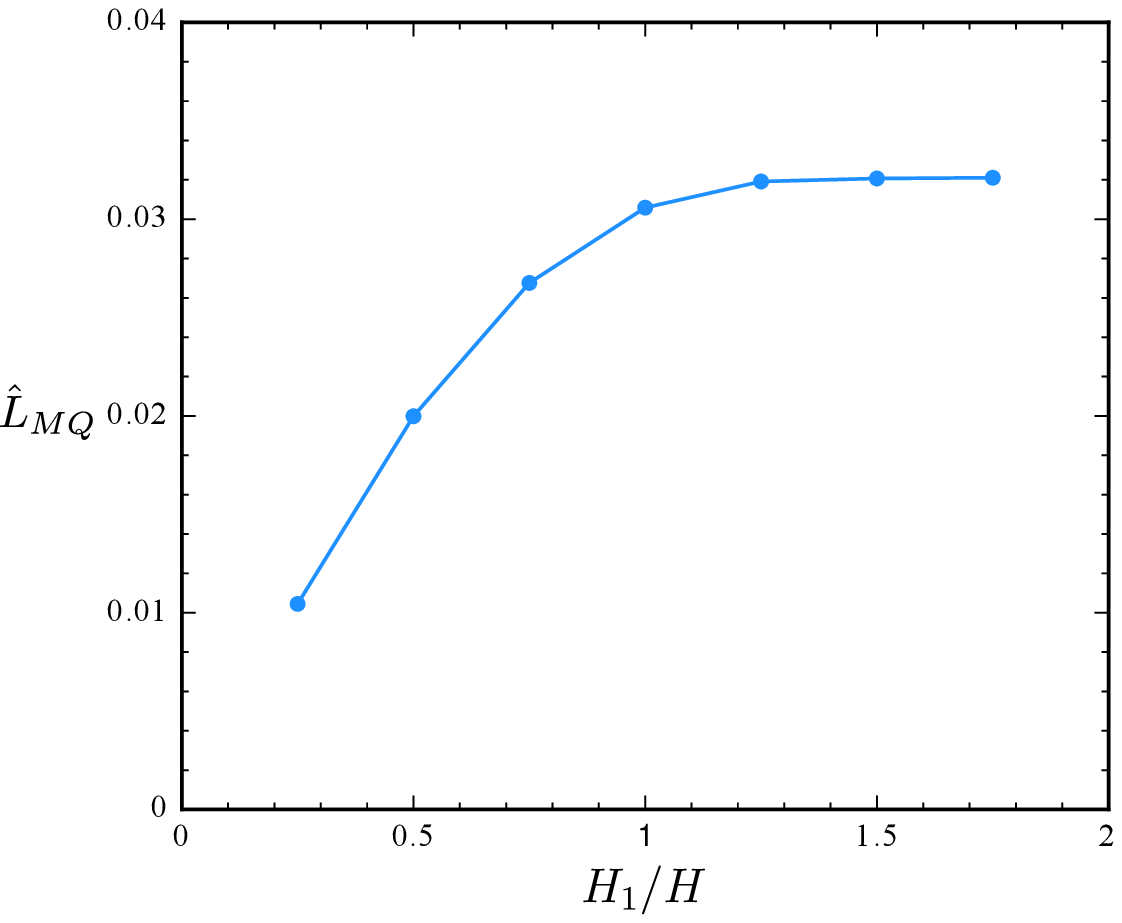}
        \label{fg:coupling_ratchet_optimize_height}
    }
    \caption{${\hat L}_{MQ}$ in the diffuse-specular configuration.
        (a) ${\hat L}_{MQ}$ plotted as a function of $L/H$ for
        $\tan\alpha=1.25$, $H_1/H=1$, and ${\rm Kn}=0.28$.
        (b) ${\hat L}_{MQ}$ plotted as a function of $H_1/H$ for
        $\tan\alpha=1.25$, $L/H=2.5$, and ${\rm Kn}=0.28$.
    }
    \label{fg:coupling_ratchet_optimize_length_height}
\end{figure*}

W\"uger \cite{Wurger2011} and Donkov et al. \cite{Donkov2011} have proposed the
capillary with ratchet surfaces as another possible configuration for Knudsen
pump. In this section, a preliminary optimization study is provided for this
purpose. For accuracy and simplicity, only one block is used in the simulations
here and the inlet/outlet is replaced by the periodic boundary condition, which
is suitable for the computation of ${\hat L}_{MQ}$. The upper wall is still
diffusely reflective, but the lower walls now have two different configurations
(as shown in figure \ref{fg:schematic_knudsen_pump}):
\begin{itemize}
    \item All the lower walls are diffusely reflective. This is referred to as
        the diffuse configuration.
    \item The lower horizontal and vertical walls are diffusely reflective, and
        the lower tilted walls are specularly reflective. This is referred to as
        the diffuse-specular configuration, the same as that adopted in Ref.
        \cite{Donkov2011}.
\end{itemize}

Technically, these two configurations are compatible with the periodic boundary
condition applied here, under which only the thermo-osmotic effect is attainable
and the coefficient ${\hat L}_{MQ}$ can be computed. As to the configuration
shown in figure \ref{fg:schematic_coupling_ratchet_less}, it is suitable for the
application of pressure difference, under which the mechano-caloric effect
coexists with the thermo-osmotic effect and hence ${\hat L}_{QM}$ and ${\hat
L}_{MQ}$ can be computed simultaneously to verify the reciprocal symmetry.

Figure \ref{fg:coupling_ratchet_optimize_alpha} shows ${\hat L}_{MQ}$ as a
function of $\tan\alpha$ for $L/H=2$, $H_1/H=1$, and ${\rm Kn}=0.28$. The left
panel in the figure is for the diffuse configuration and the right panel is for
the diffuse-specular configuration. In the diffuse configuration, the thermally
induced flows arise on both sides of the sharp edge and the net flow is
consequently diminished \cite{Hardt2013}. This explains the very small magnitude
of ${\hat L}_{MQ}$ in figure \ref{fg:coupling_ratchet_optimize_alpha_diffuse}.
In fact, if $\alpha=0$, then there will be no net mass flow in the $x$ direction
because two identical vortices are formed between the `needles' as shown in
figure \ref{fg:coupling_ratchet_field_diffuse}. The optimal value of $\alpha$
(at which ${\hat L}_{MQ}$ is maximized) is approximately given by $\tan\alpha =
1.25$. In the diffuse-specular configuration, the thermally induced flow occurs
only on the diffusely reflective surfaces, and therefore the net mass flow is
much higher than that in the diffuse configuration. The temperature distribution
and streamlines of a typical diffuse-specular configuration are shown in figure
\ref{fg:coupling_ratchet_field_specular}, and they are similar to the
corresponding results in Ref. \cite{Donkov2011}. In this case, figure
\ref{fg:coupling_ratchet_optimize_alpha_specular} shows ${\hat L}_{MQ}$ to be an
increasing function of $\tan\alpha$. For $L/H$ and $H_1/H$ used here,
$\tan\alpha$ varies between $0$ and $2$. At a larger $\alpha$, the flow explores
a smaller region between the vertical and tilted walls, with the streamlines
showing less change in direction. This means the flow is less dissipated and the
thermo-osmotic effect is stronger.

Now we focus on the diffuse-specular configuration. Figure
\ref{fg:coupling_ratchet_optimize_length} shows ${\hat L}_{MQ}$ as a function of
$L/H$ for $\tan\alpha=1.25$, $H_1/H=1$, and ${\rm Kn}=0.28$. It is seen that
${\hat L}_{MQ}$ is maximized at $L/H\approx 2.25$. Here $L/H$ can increase from
$1.25$ to $\infty$. When $L/H$ is close to $1.25$, the flow changes greatly in
direction at the lower left corner, and hence the density of dissipation is
large. On the other hand, when $L/H$ is very large, the total dissipation is also
large because of the large space involved. As a consequence, an optimal value of
$L/H$ is expected at which the total dissipation is minimized and ${\hat
L}_{MQ}$ is maximized. Figure \ref{fg:coupling_ratchet_optimize_height} shows
the ${\hat L}_{MQ}$ as a function of $H_1/H$ for $\tan\alpha=1.25$, $L/H=2.5$,
and ${\rm Kn}=0.28$. Here $H_1/H$ varies between $0$ and $2$. It is seen that
${\hat L}_{MQ}$ is an increasing function of $H_1/H$ and saturates at
$H_1/H\approx 1.25$. This is expected because the thermally induced flow is only
appreciable within a certain distance to the edge. For $H_1$ below this
distance, ${\hat L}_{MQ}$ increases with the increasing $H_1$, while for $H_1$
above this distance, ${\hat L}_{MQ}$ saturates.

\section{Concluding remarks}

The mechano-caloric effect and thermo-osmotic effect in a single-component gas
not far away from equilibrium have been investigated. The mechanisms for
micro-channels with planar surfaces and ratchet surfaces have been analyzed.
Numerical simulations have been performed to compute the off-diagonal
kinetic coefficients for the cross coupling effects as a function of the Knudsen
number. For channels with planar surfaces, our simulation results have been
compared with the S-model solution of Chernyak et al. \cite{Chernyak1979},
showing good agreement for $ 0.1 < {\rm Kn} < 10 $.
For channels with ratchet surfaces, our simulations have been performed for both
the slip and transition regimes.
The theoretical prediction for micro-channels with ratchet surfaces
in the slip regime, which gives $\overline{U}\propto{\hat L}_{MQ}\propto {\rm Kn}^C$,
has been numerically checked, showing good agreement with our simulation results.
It has also been shown that ${\hat L}_{MQ}$ and ${\hat L}_{QM}$
both exhibit their maxima at ${\rm Kn}\approx 0.28$ in the transition regime,
as anticipated theoretically.

For both types of channels, our simulation results for the off-diagonal
kinetic coefficients have confirmed Onsager's reciprocal relation.
Given the fundamental importance of reciprocal symmetry in non-equilibrium
thermodynamics, we want to point out that (i) our simulation geometry
for channels of ratchet surfaces allows the coexistence of
the mechano-caloric and thermo-osmotic effects, (ii) this coexistence
makes the numerical verification of reciprocal symmetry possible, and
(iii) this verification shows that thermodynamic consistency is ensured in
our simulation approach based on the unified gas-kinetic scheme
\cite{Xu2010,Huang2012}.
Here we remark that the reciprocal symmetry has been observed
in the slip regime and beyond. This is physically acceptable because
small pressure and temperature differences have been used in our simulations
to ensure that dynamics is slow and hence linear response is valid even if
the Knudsen number is not very small.

Since micro-channels with ratchet surfaces have the potential to be an
alternative configuration of Knudsen pump, a preliminary optimization study has
been carried out for its geometry. Two different configurations have been used
for this purpose --- (i) the diffuse configuration in which all the lower walls
are diffusely reflective, and (ii) the diffuse-specular configuration in which
the lower horizontal and vertical walls are diffusely reflective and the lower
tilted walls are specularly reflective. It turns out that the diffuse-specular
configuration leads to a much stronger thermo-osmotic effect (by an order of
magnitude). In particular, we have measured the off-diagonal coefficient ${\hat
L}_{MQ}$ as a function of various geometrical parameters, and our results can be
used to help optimize the pump design.

\section*{Acknowledgement}
This work is supported by Hong Kong RGC Grants No. HKUST604013 and C6004-14G.

\appendix
\section{Simulation Method: Unified Gas-kinetic Scheme}

The unified gas-kinetic scheme (UGKS) is a multi-scale method based on the
kinetic equation, which can be used for simulating flows of all Knudsen numbers
\cite{Xu2010,Huang2012}.

The BGK-type equation in one spatial dimension without external force is given
by
\begin{equation}
    \frac{\partial f}{\partial t}+u\frac{\partial f}{\partial x} =
    Q(f),\quad Q(f)=\frac{f^+-f}{\tau},
\end{equation}
where $f(x,t,{\bf u})$ is the velocity distribution function at $(x,t)$, ${\bf
u}=(u,v,w)$ is the particle velocity, $f^+$ is the post-collision distribution
function, and $\tau$ is the relaxation time. In the finite-volume framework, the
evolution of the distribution function and conservative variables in the $i$-th
cell are given by
\begin{equation}
    \begin{aligned}
        f_i^{n+1}=f_i^n-&
        \frac{1}{\Delta x}\int_{t^n}^{t^{n+1}}u(f_{i+1/2}-f_{i-1/2})dt \\
        +&
        \frac{\Delta t}{2}(Q_i^n+Q_i^{n+1}),
    \end{aligned}
\end{equation}
and
\begin{equation}
    {\bf W}_i^{n+1}={\bf W}_i^n-\frac{1}{\Delta x}
    \int_{t^n}^{t^{n+1}}\int\psi u(f_{i+1/2}-f_{i-1/2})d{\bf u}dt,
\end{equation}
where ${\bf W}=(\rho,\rho U,\rho E)^T$, $\psi=(1,u,{\bf u}^2/2)^T$, $U$ is the
macroscopic velocity, $E$ is the total energy density, and $d{\bf u}=dudvdw$.

The construction of the interface flux is the key to UGKS. Assuming an interface
is located at $x=0$, the accurate time evolution of the distribution function at
the interface from $t^n=0$ to $t^{n+1}=\Delta t$ is described by the solution of
the BGK-type equation along the characteristics,
\begin{equation}
    \begin{aligned}
        f(0,t,{\bf u})=&
        \frac{1}{\tau}\int_0^t f^+(-u(t-t'),t',{\bf u})e^{-(t-t')/\tau}dt' \\
        +&
        e^{-t/\tau}f_0(-ut,{\bf u}),
    \end{aligned}
\end{equation}
where $f_0$ is the initial distribution function at $t=0$. Here $f_0$ is assumed
to be linearly distributed within each cell and discontinuous at the interface,
given by
\begin{equation}
    f_0=(f_0^L+xf_{x}^L)(1-H[x])+
    (f_0^R+xf_{x}^R)H[x],
\end{equation}
where $f_0^L$ and $f_0^R$ are the reconstructed distribution functions at both
sides of the cell interface, $f_x^L$ and $f_x^R$ are the corresponding
derivatives, and $H[x]$ is the Heaviside function. The post-collision term $f^+$
is approximated by the first-order Taylor expansion on both sides of the
interface, given by
\begin{equation}
    f^+=f^+_0+g_0\left[(1-H[x])a^L x+H[x]a^R x+At\right],
\end{equation}
where $g_0$ is the Maxwell distribution which is uniquely determined by ${\bf
W}_0$,
\begin{equation}
    {\bf W}_0=\int(f_0^L H[u]+f_0^R(1-H[u]))\psi d{\bf u}.
\end{equation}
The coefficients $a^L$, $a^R$, and $A$ are computed via the partial derivatives
of the conservative variables at $(x,t)=(0,0)$, for example,
\begin{equation}
    a^L=
    \frac{1}{g_0}\left(\frac{\partial g_0}{\partial {\bf W}_0}\right)
    {\bf W}_x^L\approx
    \frac{1}{g_0}\left(\frac{\partial g_0}{\partial {\bf W}_0}\right)
    \frac{{\bf W}^L-{\bf W}_0}{x^L},
\end{equation}
where ${\bf W}^L$ is the conservative variables at the left cell, and $x^L$ is
the coordinate of the left cell center. The time derivative part $A$ is computed
via
\begin{equation}
    {\bf W}_t=-\int (a^LH[u]+a^R(1-H[u]))ug_0\psi d{\bf u}.
\end{equation}

In the present work, $f^+$ takes the form from the BGK-Shakhov model
\cite{Shakhov1972} to result in a realistic Prandtl number and the relaxation
time is $\tau=\mu/p$.

\bibliographystyle{apsrev4-1}
\bibliography{cross-coupling}
\end{document}